\documentstyle[manuscript, epsf, aps]{revtex}

\begin{document}
\title{On the nonadiabatic  geometric quantum gates }
\author{Wang Xiang-Bin\footnote{correspondence author, email: wang@qci.jst.go.jp}
 \hskip 0.3cm and  \hskip 0.2cm Matsumoto Keiji\footnote{email:keiji@qci.jst.go.jp}\\
         Imai Quantum Computation and Information Project,\\ ERATO,
Japan Sci. and Tech. Corp.\\Dani Hongo White Bldg. 201, 5-28-3, 
Hongo Bunkyo, Tokyo 113-0033, Japan}
\maketitle 
\begin{abstract}{Motivated for the fault tolerant quantum computation, a quantum gate
by adiabatic geometric phase shift is extensively investigated recently. 
Due to the enviromental noise, the decoherence time for a quantum state 
can be very short. In the adiabatic quantum computation, if we change the Hamiltonian
slowly, the decoherence error is increased, if we change the Hamiltonian faster, 
the adiabatic condition is broken and the error caused by the nonadiabaticity is increased. 
In this paper, we give
a nonadiabatic scheme for the geometric phase shift and conditional geometric phase shift
with the nuclear magnetic resonance system(NMR). 
Essentially, the new scheme is simply to add an appropriate additional field. With this additional
field, the state evolution can be controlled exactly on a dynamical phase free path.
Geometric
quantum gates for single qubit and the controlled NOT gate for two qubits are given.
In our non-adiabatic scheme, the geometric quantum 
gates can run as fast as the usual quantum gates.    }  
\end{abstract}
\section{Introduction}
It has been shown that, a quantum computer, if available, can perform certain tasks much more
efficiently than a classical Turing machine.
A fault-tolerant quantum logic gate\cite{knill1} is the central issue in realizing  the basic
constituents of a quantum information processor. 
Quantum computation via the controlled geometric phase shift\cite{ekert} provides 
a nice scenario to this purpose.
Due to its geometric property, 
a geometric phase\cite{berry,aha} shift can be 
robust with respect to certain types of operational errors. In particular, 
suppose that the spin(qubit) undergoes  a random 
fluctuation about its path in the evolution. 
The final value of the geometric phase shift 
will not be affected provided that the random fluctuation does not change
the total area on Bloch sphere\cite{ekert}. 

Recently, it was reported\cite{ekert,nature,quant,zoller,aver,los} that the conditional 
Berry phase( adiabatic cyclic geometric phase) shift can be used in  quantum computation. 
In particular, in ref\cite{ekert},
 an experiment was done
with nuclear magnetic resonance(NMR)\cite{cory,gers,jones,jones1,book} under the adiabatic condition.   

Demand on both running speed and precision of every  gate in a quantum computer 
is quite high. By the available  technology at present( and at  near future),
due to the enviromental noise, distortion to a quantum state increases 
fast with time. This requires all quantum gates  complete the task 
as fast as possible.  
However, the schemes raised previously\cite{ekert,nature} required the
adiabatic condition. That is to say, if we run the gate too fast(i.e. change the Hamiltonian rapidly), 
the distortion from the nonadiabaticity must be significant.
It has been reported in the recent experiment that\cite{ekert}, due to the 
requirement of 
adiabatic condition,
 faster running speed 
causes severe distortions to the results\cite{ekert}. The adiabatic condition
makes the fast speed and high precision conflict each other in the 
conditional Berry phase shift gate.
  Increasing the running speed of the logic gate can exponentially increase the power
of a quantum computation such as quantum factorization\cite{shor}. 
It seems that the geometric phase shift gate will  only be   practical if
it can run at a speed comparable to that of the usual quantum gate.
Therefore one is tempted to find an easy  way to make the
 conditional geometrical phase shift
through  {\em non}adiabatic state evolution. In this paper, we present an
easy  scheme to make the geometric phase shift $non$adiabatically
with NMR. In section II we study the state evolution on a cone. We give a simple way to
nonadiabatically control the state evolution on the cone. With an appropriate setting, the
state evoleves on a dynamical phase free path. In section III we implement the result of
section II to the NMR system. A scheme for nonadiabatic conditional geometric shift is given.
In section IV we investigate the nonadiabatic geometric
realization of various types of quantum logic gates, including the phase shift gate, the Hadamad gate
and the NOT gate for single qubit, and the C-NOT( controlled-NOT) gate for two qubits. We give a concluding remark in section V. 
 \\
\section{ Exact state evolution  on the cone.}
It is well known that a spin half nucleus can gain a geometric
phase shift in the conical evolution.
We now assume that the initial spin state is on the cone and we will demonstrate a method to 
 make it evolve nonadiabatically.
The Hamiltonian for a spin  in a constantly rotating magnetic field
\footnote{We shall extend it to the case of arbitrary time-dependent rotating speed in the end of this section.}  
is
\begin{eqnarray}\label{e1}
H(t)=\left[\omega_0\sigma_z +\omega_1 \sigma_x(t)\right]/2
.\end{eqnarray}
Here $\sigma_x(t)=\left(\begin{array}{cc}
0&e^{-i\gamma t}\\e^{i\gamma t}
&0\end{array}\right)$, 
$\omega_0$ is the amplitude of vertical field and $\omega_1$ is the amplitude of the horizontal field, which
is rotating around $z-$axis in  the constant  angular speed $\gamma$.
The initial state  $|\psi_0\rangle$ is an
eigenstate of $H_0=H(0)$. 
Explicitly,  
\begin{eqnarray}
|\psi_0\rangle=\cos\frac{\theta}{2}|\uparrow\rangle+\sin\frac{\theta}{2}|\downarrow\rangle
\end{eqnarray} and
\begin{eqnarray}
\cos\theta=\omega_0/\sqrt{\omega_0^2+\omega_1^2}.\end{eqnarray}
To know the cause of the state distortion, we study the time evolution first.(The similar idea has been used
for the adiabatic rotational splitting with NQR previously\cite{robert}.)
The time evolution operator $U$ generated by the Hamiltonian $H_0$ is determined by
the time-dependent Schrodinger equation
\begin{eqnarray}
i\frac{\partial}{\partial t}U=H(t) U.
\end{eqnarray}
Solving this equation we obtain
\begin{eqnarray}
U(t)=e^{-i\gamma\sigma_z t}e^{-i(H_0-\gamma \sigma_z/2)t}.
\end{eqnarray}
In particular, at time $t=\tau=2\pi/\gamma$, when the external field completes a $2\pi$ rotation, the state is 
evolved to
\begin{eqnarray}
\psi(\tau)=e^{-i\pi-i(H_0-\gamma \sigma_z/2)\tau}|\psi_0\rangle.
\end{eqnarray}
The adiabatic approximation assumes that
at time $\tau$, the state completes a cyclic evolution provided that $\gamma$ is small. 
However, due to the decoherence time limitation, $\gamma$ cannot be too small. 
The non-zero $\gamma$ distorts the state in the evolution and 
makes it {\em noncyclic} at time $\tau$ .   
This non-zero $\gamma$ in the adiabatic approximation causes two types of errors. One is that
the noncyclic completion of the evolution at time $\tau$ will cause further errors in the succeeding operations in the sequence.
(A sequence of operations is used 
to remove the dynamic phase in realizing the geometric quantum gate\cite{ekert}.)
The other is that the geometric phase acquired over period $\tau$ is not $-\pi(1-\cos\theta)$, 
as for the ideal adiabatic cyclic evolution. 

Now we give an easy way to exactly control 
 the state evolution on the cone.  
Here the external field can be  rotated around $z-$axis arbitrarily fast. We use $|\psi_0\rangle$,
the eigenstate of $H_0$ for the initial state. 
We  switch on a static vertical magnetic field 
$\omega_z=\gamma$ while the external field is rotated around $z-$axis.
In this way, the new time-dependent Hamiltonian is 
\begin{eqnarray}
H_W(t)=\frac{1}{2}\left[(\omega_0+\gamma) 
\sigma_z+\omega_1 \sigma_x(t)\right].
\end{eqnarray} The 
time evolution operator $U_W(t)$ generated by this Hamiltonian 
satisfies the Schrodinger equation
\begin{eqnarray}\label{sch}
i\frac{\partial}{\partial t}U_W(t)=H_W(t)U_W(t)
\end{eqnarray}
with the boundary
condition $U_W(0)=1$. The state  at time $t$ is related to the state
at time $0$ by $|\psi(t)\rangle=U_W(t)|\psi_0\rangle$, $\psi_0=\psi(0)$.
 Denoting $R=e^{i \gamma t\sigma_z /2}$ we obtain the following equivalent equation
\begin{eqnarray}
i\partial(RU_W)/\partial t=H_0(RU_W).
\end{eqnarray}
Now that $H_0$ is time-independent  we have 
\begin{eqnarray}
RU_W(t)=e^{-iH_0t}.
\end{eqnarray} 
This is equivalent to
\begin{eqnarray}
U_W(t)=e^{-i\gamma t\sigma_z/2}e^{-iH_0 t}.
\end{eqnarray}
  Consequently, at any time $t$, state 
\begin{eqnarray}|\psi(t)\rangle=e^{-i\lambda t}
e^{-i\gamma t\sigma_z /2 }|\psi_0\rangle\end{eqnarray} is
exactly the instantaneous eigenstate of Hamiltonian $H(t)$, where
$\lambda$ is the eigenvalue of $H_0$ for eigenstate $|\psi_0\rangle$,
\begin{eqnarray}
\lambda=\pm\sqrt{\omega_0^2+\omega_1^2}.
\end{eqnarray}
 In particular, at time $\tau$, 
\begin{eqnarray}
|\psi(\tau)\rangle=e^{-i\pi-i\lambda \tau}|\psi_0\rangle.
\end{eqnarray} 
It   only differs to $|\psi_0\rangle$ by a phase factor.
Note that the time evolution operator $U_W(t)$ here is generated by the Hamiltonian 
$H_W(t)$ instead of $H(t)$. 

Here
the additional vertical field $\gamma$ plays an important role. Without this field, the time
evolution operator generated by $H(t)$ is
\begin{eqnarray}
U(t)=e^{-i\gamma t\sigma_z/2}e^{-iH_1 t}
\end{eqnarray}
and $H_1=H_0-\gamma\sigma_z/2$. 
Obviously, this time evolution operator $U(t)$ will distort a qubit with the initial
state $|\psi_0\rangle$ in the evolution. But this $U(t)$ can exactly
control the qubit with the initial state
$|\psi_1\rangle$, the eigenstate of $H_1$. If initially we set the
spin state to $|\psi_1\rangle$, then the spin will evolve exactly on its cone without the 
additional field
$\gamma$.  So,
we have two ways to control the spin evolution exactly. 
We can use the additional field $\gamma$, if the initial state is set to be $|\psi_0\rangle$.
Alternatively, we can also set the initial state to be $|\psi_1\rangle$ and 
then we need not add
any additional field when the external field is rotated.
In this letter, we adopt the former one, i.e. we set the initial spin state to $|\psi_0\rangle$.

The additional magnetic field $\gamma$ can make the
 state $|\psi_0>$ keep up the rotating field exactly in the evolution. So the state completes a cyclic
evolution when the field completes a $2\pi$ rotation.
However, after a cyclic evolution, the total phase shift for the state in general
includes both dynamic and geometric 
contribution\cite{piller}. It is possible to restrict the state  evolution on the dynamic phase path if we  choose certain specific value of the rotating speed
$\gamma$( and also the additional field $\gamma$). For this purpose, we require
\begin{eqnarray}\label{restr}
\gamma\cos\theta=-\sqrt{\omega_0^2+\omega_1^2}.
\end{eqnarray}
The solution is
\begin{eqnarray}\gamma=-\frac{\omega_1^2+\omega_0^2}{\omega_0}.
\end{eqnarray}
The negative sign in the right hand side indicates that the additional field $\gamma$
is anti-parallel with the field $\omega_0$.
By this setting, the total magnetic field in $H_W$ is always 
"perpendicular" to the state vector expressed in the Bloch ball. Details of this are shown in Fig. 1. 
One can easily show that the instantaneous dynamical phase
\begin{eqnarray}
<\psi_0|U_W^\dagger(t)H_W(t)U_W(t)|\psi_0>=0.
\end{eqnarray}

We have assumed above that the field rotates at a constant speed.
Actually, we can easily modify the  above scheme for 
arbitrary time-dependent rotating speed $\gamma_a(t)$, 
 $\int^\tau_0 \gamma_a(t)dt=2\pi$. In this case, we need only change the term $\gamma t$
in $H_W(t)$, $U_W(t)$, $R$ and $\psi(t)$ into $\int^t_0\gamma_a(t')dt'$ accordingly.
Consequently, the the additional vertical field is now a time-dependent 
field $\omega_z(t)=\gamma_a(t)$ instead of a static field. To remove the 
dynamical phase,
we can use the time-dependent magnetic field $\omega_1(t)$ and $\omega_0(t)$.
We require
\begin{eqnarray}
\gamma(t)=-\frac{\omega_1^2(t)+\omega_0^2(t)}{\omega_1(t)}
\end{eqnarray}
and
\begin{eqnarray}
\omega_1(t)/\omega_0(t)=\omega_1(0)/\omega_0(0)=\tan\theta
\end{eqnarray}

This extension to the time-dependent case can be important
in case it is difficult to rotate the field (or the fictitious field\cite{nature}) in a
 constant angular speed. Punctual results can be obtained
here through the exact feedback system where the value of additional vertical field is always
instantaneously equal to the  angular velocity of the rotating field. 

Thus we see, by adding an additional magnetic field that is equal to the rotating
frequency of the external field, we do get the {\it exact} result. 
 Obviously if we
rotate the field  inversely, additional vertical field  should be 
in the inverse direction($-z$) accordingly. 

\section{ The nonadiabatic conditional geometric phase shift.} 
\subsection{ NMR system and the rotational framework.} 
Consider the interacting nucleus spin pair(spin $a$ and spin $b$) in the NMR 
quantum computation\cite{cory,gers,jones,jones1}.
If there is no horizontal field the Hamiltonian for the two qubit system is
\begin{eqnarray}
H_i=\frac{1}{2}(\omega_{a} \sigma_{za}+\omega_b\sigma_{zb}
+ J \sigma_{za}\cdot\sigma_{zb}) \end{eqnarray},
where $\omega_{a}(\omega_b)$ is the resonance frequency for spin $a(b)$ 
 in a very strong 
static magnetic field(e.g. $\omega_a$ can be $500$MHz\cite{ekert}) ,  $J$ is the interacting constant between
nuclei and $\sigma_{za}=\sigma_{zb}=\sigma_z=\left(\begin{array}{cc}
1&0\\0&-1\end{array}\right)$.  
After adding a circularly polarized RF field in horizontal
plane,
the Hamiltonian for spin $a$ in static framework is
\begin{eqnarray}
H'=\frac{1}{2}\omega_0\sigma_z+\frac{1}{2}\omega_a'\sigma_z+\frac{1}{2}\omega_1
\left(\begin{array}{cc}0&e^{-i\omega_a't}\\e^{i\omega_a't}&0\end{array}\right).
\end{eqnarray}
Here $\omega_a'$ and $\omega_1$ are the angular frequency and amplitude of the RF field respectively,
\begin{eqnarray}
\omega_0=\omega_a-\omega_a'\pm J
\end{eqnarray} and the $"\pm"$  sign in front of $J$ is dependent on 
the specific state of spin $b$, up or down respectively. To selectively manipulate spin $a$, we shall 
use the rotational framework that is rotating around $z-$axis
in speed $\omega_a'$.
We assume $\omega_a'$  close to $\omega_a$ but quite different from $\omega_b$.
The Hamiltonian for spin $a$ in the rotational framework is   
\begin{eqnarray}
H_a=R'H'R'^{-1}+i(\partial R'/\partial t)R'^{-1}=\frac{1}{2}\omega_0\sigma_z+
\frac{1}{2}\omega_1\sigma_x(t)=H(t)
\end{eqnarray}
and $R'=e^{i\omega_a'\sigma_zt/2}$. Note here $H_0$ is dependent on  state of spin $b$ 
through $\omega_0$.
In the rotational framework, if the horizontal field is rotated in the  angular speed
$\gamma$, 
the Hamiltonian for spin $a$ is just $H(t)$, as defined in eq(\ref{e1}). In the NMR system, we require $|\psi_0\rangle$ 
to be the eigenstate
of $H_0$ in rotational the framework.  Previously\cite{ekert}, state 
$|\psi_0\rangle$ for spin $a$ was produced  adiabatically. 
In the following we demonstrate how to 
nonadiabatically produce the state $|\psi_0\rangle$.
\subsection{ Creating the conditional initial state with NMR.}
In rotating the state around $z-$axis, we have started from state $|\psi_0\rangle$, 
which is the eigenstate
of $H_0$. State $|\psi_0\rangle$ can be created from the spin state $|\uparrow\rangle$ or $|\downarrow\rangle$. 
We have  a $non$adiabatic way to  
create state $|\psi_0\rangle$ for spin $a$.  
Note that we are working in the rotational framework. We denote 
$\delta=\omega_a-\omega_a'$.
We can use the following sequence of operations
to create the conditional angle $\theta$.(For simplicity we will call the following sequence as $S$ operation later on.)
We have 
\begin{eqnarray}\label{scheme}
S=\left[\frac{\pi}{2}\right]^y\rightarrow
J'(\varphi_{\pm}(t_c))
\rightarrow\left[-\delta \cdot t_c\right]^z
\rightarrow \left[\frac{\pi}{2}\right]^x
\rightarrow\left[{-\varphi'}
\right]^y
\end{eqnarray}
Here all terms inside the $[\cdots]$ represent the Bloch sphere
rotation angles caused by  RF pulses. 
The superscripts indicate the axis the Bloch sphere is rotated around. $J'( \varphi_{\pm}(t_c))$ 
is the time evolution over period $t_c$ by the 
Hamiltonian $\frac{1}{2}(\delta\pm J)\sigma_z$.
 This evolution rotates spin $a$ around $z-$axis for an angle
 $\varphi_{\pm}=(\delta\pm J)t_c$.
(Or $\varphi_{\pm}+\pi$, if spin $a$ is down initially. 
For clarity we omit this case and alway assume spin $a$ is initially up.) 
After this $S$ operation, the angle between spin $a$ and the $z$-axis is
$\theta_{\pm}=\frac{\pi}{2}-(\varphi'+\varphi_{\pm})$(see Fig. 2). 
To ensure the state to be eigenstate of Hamiltonian $H_0$
after $S$ operation 
we require $ \tan(\varphi'+Jt_c)=\frac{\delta+J}{\omega_1}$ and $\tan (\varphi'-Jt_c/2)=\frac{\delta-J}{\omega_1}$ simultaneously. This is equivalent to 
\begin{eqnarray}\label{const}
\left\{\begin{array}{c}Jt_c=
(\arctan\frac{\delta+J}{\omega_1}-\arctan\frac{\delta-J}{\omega_1})/2\\
\varphi'=(\arctan\frac{\delta+J}{\omega_1}+\arctan\frac{\delta-J}{\omega_1})/2\end{array}
\right. .\end{eqnarray}
From this constraint, given specific values of $\delta$, 
$J$ and $\omega_1$, we can easily obtain the scaled time $J\cdot t_c$ (Fig. 3)
and the angle $\varphi'$(Fig. 4) in controlling the $S$ operation. In particular, for $\delta/J=1.058$,  the value adopted
in the recent NMR experiment, the relation curves for $Jt_c$ vs $\omega_1$ and $\varphi'$ vs 
$\omega_1$ are shown 
in Fig. 5.  
\subsection{ The nonadiabatic conditional geometric phase shift.} 
After the $S$ operation, the state of spin $a$ 
is an eigenstate of Hamiltonian
$H_0$ no matter whether spin $b$ is up or down. 
After this $S$ operation we can use the additional magnetic field  to nonadiabatically
control  the state evolution on the cone. However, here we need again remove the dynamical phase
contribution. Note the situation here is different from the single qubit case.
Qubit $b$ could be either up or down. We need choose the appropriate $\gamma$ value so that
the instantaneous dynamic phase for qubit $a$ is always zero no matter qubit $b$ is up or down.
So in this case  the following simultaneous conditions are required.
\begin{eqnarray}
\gamma\cos\theta_{\pm}=-\sqrt{(\delta\pm J)^2+\omega_1^2}.
\end{eqnarray}
Setting
\begin{eqnarray}
\omega_1=\sqrt{\delta^2-J^2}.
\end{eqnarray}
will make the simultaneous conditions hold. 
With this setting
 we propose the following 
scheme
to make the $non$adiabatic conditional geometric phase shift.    
\begin{eqnarray}
S\rightarrow\left(\begin{array}{c}\gamma\\{C}\end{array}\right)
\rightarrow S^{-1}
.\end{eqnarray}
The term $\left(\begin{array}{c} {\gamma}\\{ C}\end{array}\right)$  
 represents  
doing operation $C$  with an additional
vertical field $\gamma$. 
$C$ represents
 rotating the external field around $z-$axis for $2\pi$ in a uniform speed $\gamma$. 
We should choose the rotating direction so that additional field $\gamma$
is anti-parallel the field $\delta\pm J$, ($\delta>J$). 
Since qubit $a$ evolves on the dynamical phase free path,
the scheme raised here can remove the dynamic phase and
retain only the geometric phase. 
It can be shown that the geometric phase
acquired after the operation is $\Gamma_+$, $-\Gamma_+$, $\Gamma_-$ and 
$-\Gamma_-$ respectively for the four
different initial state
($|\uparrow\uparrow\rangle,|\uparrow\downarrow\rangle,
|\downarrow\uparrow\rangle,|\downarrow\downarrow\rangle$), 
\begin{eqnarray}
\Gamma_\pm=-\pi- \pi \cos\theta_\pm
\end{eqnarray}
and\begin{eqnarray}
\cos\theta_\pm=\frac{\delta\pm J}
{\sqrt{(\delta\pm J)^2+\omega_1^2}}=\sqrt{\frac{\delta\pm J}{2\delta}}\end{eqnarray}. 
\section{Geometric quantum logic  gates} 
\subsection{Geometric gates for single qubit}
To make a quantum computation, we need both single qubit gates and the two qubit C-NOT gate.
It is impossible to make any single qubit geometric gate with NMR in the adiabatic condition
because the interaction among different qubits cannot be switched off. Though the interaction
 is small,
the effects from the interaction may accumulate significantly after a long term.
Strictly speaking, the adiabatic 
geometric quantum computation raised previously\cite{ekert}
 is only a partial geometric quantum computation.
However, in the our non-adiabatic scheme, we can easily make single qubit gate. 
In this nonadiabatic case, the
effect of the interaction among qubits is negnigible because the operation time is very short.

In section II,  we have required the initial state of the qubit be the eigenstate
of the initial Hamiltonian, $H_0$. 
However, this requirement is not necessary
in making the geometric quantum gates, instead of  making
 geometric phase shift to the qubit directly. If we are only interested
to the geometric quantum gates instead of making the phase shift
to the qubit itself, we can directly use arbitrary initial state. 
We now demonstrate this by some examples. Without any loss of generality, we
suppose the initial state of the single qubit  is $|\uparrow>$ or $|\downarrow>$.
 We start from the phase shift gate.
We set the initial magnetic field ${\bf \omega}$ in the $z$ direction. We then rotate $\omega$
anti-clockwise around $z'$ axis. The rotating angular velocity is $\gamma$.
 The angle between $z'$ and $z$ axis is $\theta_0$( see Fig. 6). Meanwhile
we add the additional field $\gamma$. The rotating speed and additional field
should satisfy eq.(\ref{restr}). Note that in this case, the additional
field is in the  $-z'$ direction. After a $2\pi$ angle rotation the initial states will gain 
phase shifts according to the following equation
\begin{eqnarray}
\left(\begin{array}{c}|\uparrow>\\|\downarrow>\end{array}\right)
=W_p(\theta_0)\left(\begin{array}{c}|\uparrow>\\|\downarrow>\end{array}\right)
\end{eqnarray}
and 
\begin{eqnarray}
W_p(\theta_0)=\left(\begin{array}{cc}
e^{-2\pi i\cos \theta_0} & 0\\0& e^{2\pi i \cos\theta_0}
\end{array}\right).
\end{eqnarray}
The phase shift here is determined by area enclosed by the state evolution loop on Bloch sphere, i.e.
by angle $\theta_0$. Since $\theta_0$ can take arbitrary value, we can
make arbitrary phase shift gate through the above equation.

Now we consider the Hadamad gate, which is another fundamentally important 
 gate in quantum computation.
We come back to  $z$ axis.  
Now $\theta_0$ is the angle between magnetic field $\bf{\omega}$  and $z$ axis. With the 
constraint of eq(\ref{restr}), suppose the additional magnetic field and 
rotating speed of $\omega$ is $\gamma$. State $|\uparrow>$ or $|\downarrow>$
can be written  in the following
linear superposed forms respectively
\begin{eqnarray}
|\uparrow>=\cos\frac{\theta_0}{2}|+>-\sin\frac{\theta_0}{2}|->
\end{eqnarray}
and
\begin{eqnarray}
|\downarrow>=\sin\frac{\theta_0}{2}|+>+\cos\frac{\theta_0}{2}|->
\end{eqnarray}  
  where the states $|\pm>$ are the spin states parallel or anti-parallel to the direction of ${\bf \omega}=\sqrt{\omega_0^2+\omega_1^2}$.
After the field rotates one loop around $z$ axis, the initial state $|\uparrow>$ or
$|\downarrow>$
changes by the following formula
\begin{eqnarray}
\left(\begin{array}{c}|\uparrow>\\|\downarrow>\end{array}\right)
=W_s\left(\begin{array}{c}|\uparrow>\\|\downarrow>\end{array}\right)
\end{eqnarray}
 where 
\begin{eqnarray}
W_s=\left(\begin{array}{cc}\cos\frac{\theta_0}{2}&-\sin\frac{\theta_0}{2}\\
\sin\frac{\theta_0}{2}&\cos\frac{\theta_0}{2}\end{array}\right)\left(\begin{array}{cc}
e^{i\Gamma}& 0\\ 0 & e^{-i\Gamma}\end{array}\right)
\left(\begin{array}{cc}\cos\frac{\theta_0}{2}&\sin\frac{\theta_0}{2}\\
-\sin\frac{\theta_0}{2}&\cos\frac{\theta_0}{2}
\end{array}\right)\end{eqnarray}
i.e.
\begin{eqnarray}
W_s=\left(\begin{array}{cc}\cos\Gamma\cos\theta_0+i\sin\Gamma\sin\theta_0
& i\sin\Gamma\sin\theta_0\\ i\sin\Gamma\sin\theta_0& \cos\Gamma-i\sin\gamma\sin\theta_0
\end{array}\right).
\end{eqnarray}
Here $\Gamma$ is the geometric phase gained by state $|+>$ after it completes the loop around 
$z$ axis.
If we take $|\sin\Gamma\sin\theta_0|=\frac{\sqrt 2}{2}$(root $\theta_0$ for this equation obviously
exists), this $W_s$ is
\begin{eqnarray}
W_{s0}=\frac{\sqrt 2}{2}\left(\begin{array}{cc}e^{i\phi}& i\\i&-e^{-i\phi}\end{array}\right).
\end{eqnarray}
Here $\phi=\tan^{-1}\frac{\sin\Gamma\sin\theta_0}{\cos\Gamma}$. 
Combining this $W_{s0}$ gate with an appropriate phsae shift gate we can obtain the Hadamad gate.
Explicitly we have
\begin{eqnarray}
\left(\begin{array}{cc}e^{-i(\phi-\pi/2)}&0\\0&1\end{array}\right) W_{s0}
\left(\begin{array}{cc}e^{-i(\phi-\pi/2)}&0\\0&1\end{array}\right) 
=\frac{\sqrt 2}{2}\left(\begin{array}{cc}1&1\\1&-1\end{array}\right)=W_H.
\end{eqnarray}

Obviously, by combining the gate $W_H$ and the $W_p$, we can also make the NOT gate.
It is
\begin{eqnarray}
W_HW_p(\theta_0=\cos^{-1}\frac{1}{4})W_H=i\left(\begin{array}{cc}0&1\\1&0\end{array}\right).
\end{eqnarray}
 The NOT gate
of single qubit played an important role in the adiabatic scheme to cancel the dynamical 
phase\cite{ekert,nature}. 

 So far we have demonstrated the geometric realization for several types of 
logic gates in single
qubit operation. Here geometric phase is the only contribution in the time evolution. 
Actually, all single qubit gate can be realized in a similar way.
The running speed of the above
geometric gates can be in principle
arbitrarily fast.  
\subsection{Geometric C-NOT gate}
With the conditional geometric phase shift, it is very easy to construct the C-NOT gate.
To do so, we take
$\delta=\frac{4\sqrt 7}{7}J$, then we have
\begin{eqnarray}
\Gamma_+=\Gamma_--\frac{\pi}{2}
\end{eqnarray}
and
\begin{eqnarray}
S\left(\begin{array}{c} \gamma \\C\end{array}\right)S^{-1}=\left(\begin{array}{cc}
e^{i(\Gamma_--\pi/2)} &0\\0&e^{-i(\Gamma_--\pi/2)}\end{array}\right)
\end{eqnarray}  
 if qubit $b$ is up. And
\begin{eqnarray}
S\left(\begin{array}{c} \gamma \\C\end{array}\right)S^{-1}=\left(\begin{array}{cc}
e^{i\Gamma_-} &0\\0&e^{-i\Gamma_-}\end{array}\right)
\end{eqnarray}
if qubit $b$ is down. We can use   a specific phase shift gate $W_p$ on qubit $a$ to 
convert the above form of conditional
phase shift gate into
\begin{eqnarray}
S\left(\begin{array}{c} \gamma \\C\end{array}\right)S^{-1}\left(\begin{array}{cccc}
e^{-i\Gamma_-} &0&0&0\\
0&e^{i\Gamma_-}&0&0\\
0&0&e^{-i\Gamma_-}&0\\
0&0&0&e^{i\Gamma_-}
\end{array}\right)=
\left(\begin{array}{cccc}
-i &0&0&0\\
0&i&0&0\\
0&0&1&0\\
0&0&0&1
\end{array}\right).
\end{eqnarray}
Sandwitching it by two Hadamad gates to the single qubit $a$ we obtain the C-NOT gate, i.e.
\begin{eqnarray}\frac{\sqrt 2}{2}
\left(\begin{array}{cccc}1&1&0&0\\1&-1&0&0\\0&0&1&1\\0&0&1&-1\end{array}\right)
\left(\begin{array}{cccc}
-i &0&0&0\\
0&i&0&0\\
0&0&1&0\\
0&0&0&1
\end{array}\right)\frac{\sqrt 2}{2}\left(\begin{array}{cccc}1&1&0&0\\1&-1&0&0\\0&0&1&1\\0&0&1&-1\end{array}\right)
=\left(\begin{array}{cccc}
0 &-i&0&0\\
-i&0&0&0\\
0&0&1&0\\
0&0&0&1
\end{array}\right).
\end{eqnarray}

\section{ Concluding remark.} To make the conditional geometric phase shift gate 
we need be able to control two 
elementary operations. One is to exactly control the cyclic state evolution
on a cone; the other is to produce the initial state on the cone with an angle 
$\theta_{\pm}$ conditional on the other bit. 
Previously\cite{ekert}, these two elementary operations were done  
adiabatically. 
We have shown that both of these two tasks can be done nonadiabatically.
We have demonstrated the non-adiabatic   
 geometric phase shift and quantum 
logic gates for both single qubit and 
two qubit case. The single qubit geometric
gate can run arbitrarily fast in princilpe.
The running speed of two qubit gate is is only limited by the $J$,
the value of the qubits interraction. This is the case for all normal
conditional quantum gates. 
So it possible to run the geometric quattum gate in a speed
 comparable to
that of the normal quantum gate.
The idea on nonadiabatic geometric phase shift gate demonstrated by the NMR system here
should in principle  also work for the other two level systems, 
such as Josephson Junction system\cite{nature} and the harmonic oscillator system\cite{quant,wang}, 
where the decoherence time can be much shorter. It should be interesting to study the nonadiabatic
scheme for the nonabelian geometric quantum computation\cite{duan1,pacos1,pacos2,loydg}\\
{\bf Acknowledgment} We thank Prof Imai Hiroshi for supporting,
Dr Hachimori M., Mr Tokunaga Y and Miss Moriyama S for various helps. 
W.X.B thanks Prof. Ekert A(Oxford) for suggestions. W.X.B. also thanks Dr. Pan J.W.(U. Vienna), Dr Kwek LC(NIE) for discussions
 and Prof
Oh CH(NUS) for pointing out ref\cite{nature}. 
\begin{figure}
\begin{center}\epsffile{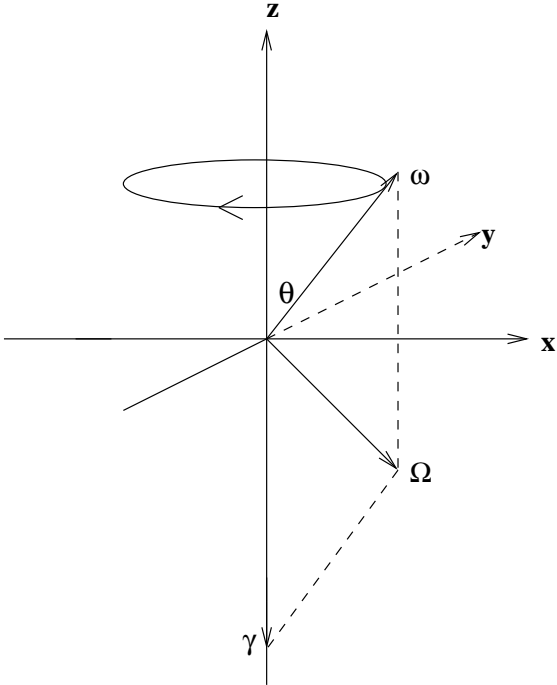}
\caption{ {\bf Control the conical evolution exactly.} 
The initial state is the eigen state of $H_0$. $\omega$ is the field in $H_0$.
$\omega$ is rotating around $z$ axis in the direction indicated by the arrow.
Additional  field $\gamma$  is equal to the rotating speed of $\omega$. With this
additional field, the state will evolve keep up the field rotating $\omega$ exactly. 
We can see that, choosing an appropriate $\gamma$, the total field is instattaneously
perpendicular to the field $\omega$, thus eigen state of $H_0$ evolves in a dynamic phase free
path.
 }\end{center}
    \end{figure}
  
\begin{figure}
\begin{center}\epsffile{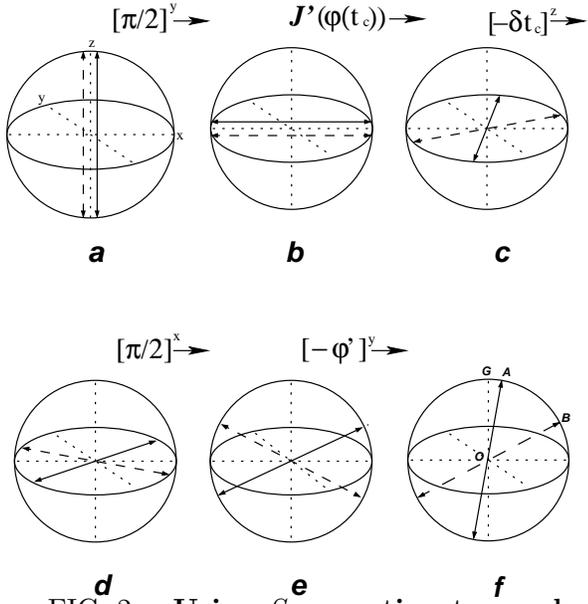}
\caption{ {\bf Using $S$ operation to produce 
the eigenstate of $H_0$ nonadiabatically.} Picture   {\sl a} shows the possible initial state
for spin $a$. The up(down) arrow on the solid line represents the up(down) state for spin $a$
while spin $b$ is up. 
The up(down) arrow on the dashed line represents the up(down) state for spin $a$
while spin $b$ is down. In picture {\sl f},  
angle $GOA$ is $\theta_+ =\arctan \frac{\omega_1}{\delta +J} $ and
angle $ GOB$ is $\theta_-=\arctan \frac{\omega_1}{\delta -J} $.
 }\end{center}
    \end{figure}
\begin{figure}
\begin{center}
\epsffile{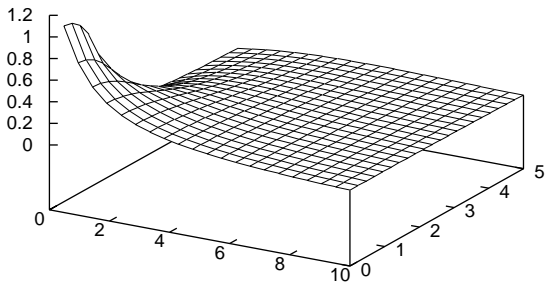}
\end{center}
\caption{{\bf The time control in $S$ operation}. 
Parameter $\omega_1/J$ varies from 0 to $10$, $\delta/J$ varies from $0$ to $5$. 
Vertical axis is for the  scaled time $Jt_c$.}
\end{figure}
\begin{figure}
\begin{center}
\epsffile{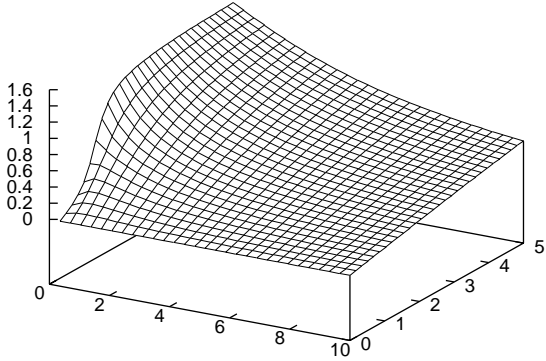}
\end{center}
\caption{{\bf The rotating angle control in $S$ operation}. 
Parameter $\omega_1/J$ varies from $0$ to $10$, $\delta/J$ varies from $0$ to $5$. 
Vertical axis is for 
the angle $\varphi'$. }
\end{figure}
\begin{figure}
\begin{center}
\epsffile{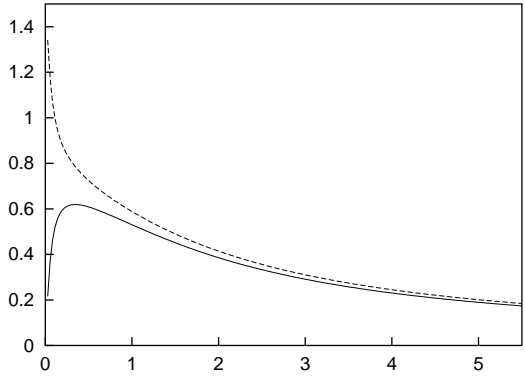}
\end{center}
\caption{ {\bf The time and rotating angle control in $S$ operation for specific $\delta/J$ value}. 
Horizontal axis represents $\omega_1/J$. 
Vertical axis is for the scaled control $Jt_c$(the solid line) or
the angle $\varphi'$(dashed line). Here $\delta/J=1.058$, as for the experimental condition[2].
By this figure, given the specific values of $\omega_1/J$, we can always find the corresponding
point in the two curves thus we can take the suitable time control(for $t_c$) and rotation
control(for $\varphi'$) in the $S$ operation.}   
\end{figure}
\begin{figure}
\begin{center}
\epsffile{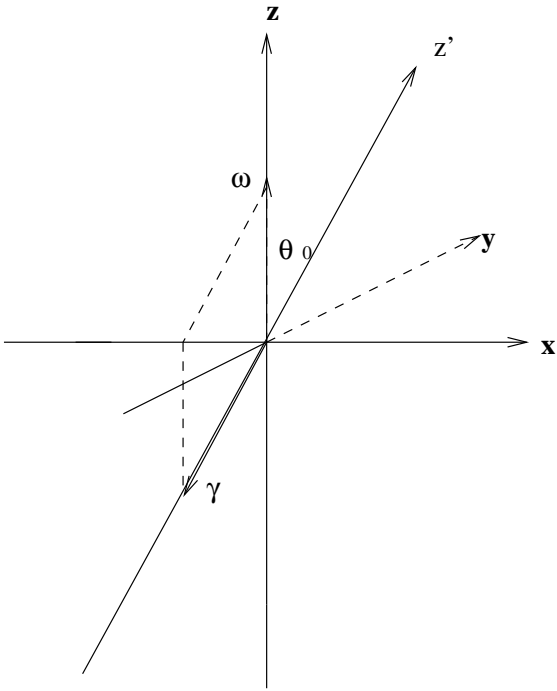}
\end{center}
\caption{{\bf Rotating the field around $z'$ axis}. 
Here the field $\omega$ is in $+z$ direction, rotates around $z'$ axis. Additional field $\gamma$
is in $-z'$ direction.}
\end{figure}

\end{document}